\documentclass[11pt]{article}

\usepackage{amsmath,amssymb,bbold,epsfig}

\begin{document}

\begin{center}
{\Large \bf 
Oscillations and leptogenesis:\\ 
\vspace*{0.2cm} what can we learn about right-handed neutrinos?
\footnote{Including material presented in the talks given at  
XV IFAE, Lecce (Italy), April 2003;
IV NANP Conference, Dubna (Russia), June 2003;
VIII TAUP Workshop, Seattle (USA), September 2003.}}
\\
\vspace{0.8cm}
{\large 
M. Frigerio \footnote{E-mail address: frigerio@physics.ucr.edu}
}
\vspace{0.2cm}

\vspace*{0.1cm}

{\em INFN and International School for Advanced Studies (SISSA/ISAS), \\
Via Beirut 2-4, 34014 Trieste, Italy.}\\ 
{\em Department of Physics, University of California, Riverside, 
CA 92521, USA.}

\end{center}

\abstract{
Inverting the type-I seesaw formula,  we reconstruct
the mass matrix of the heavy right-handed neutrinos $N_i$.
We analyze how the data on neutrino oscillations
affect the structure of this matrix.
Under the assumption of hierarchical Dirac-type neutrino masses $m_{Di}$,
we compute the mixing angles among $N_i$, their masses $M_i$
and the lepton asymmetries $\epsilon_i$ generated in their decays.
Unless special cancellations take place, one finds
$M_i\propto m_{Di}^2$ and the generated
baryon-to-photon ratio $\eta_B$ is much smaller than 
the observed value.
We show that successful baryogenesis via leptogenesis occurs in
a unique special case, which corresponds to
mass degeneracy and maximal mixing of $N_1$ and $N_2$.}


\section{Introduction}

The smallness of neutrino masses is naturally understood if neutrinos are
Majorana particles, since the Majorana mass term can originate only
from a five-dimensional operator \cite{wein1,wein2}
suppressed by a large energy scale: $LL\phi\phi / \Lambda$,
where $L$ and $\phi$ are the Standard Model (SM) lepton and Higgs doublets,
respectively.
A very simple and appealing way to generate this operator is
the type-I seesaw mechanism  \cite{yana,gla,GRS,mose}. 
In this case $\Lambda$ is the mass scale of
right-handed (RH) neutrinos, which are SM singlet Majorana particles.
The low-energy neutrino mass matrix $m$ 
is given in terms of the 
Majorana mass matrix of the RH neutrinos, $M_R$, and the Dirac 
mass matrix, $m_D$, as 
\begin{equation}
m =- m_D M_R^{-1} m_D^T ~.
\label{seesaw}
\end{equation}
While the elements of $m_D$ are expected to be at or below the 
electroweak scale ($\approx 100$ GeV), 
the characteristic mass scale of RH neutrinos is naturally the GUT or 
parity breaking scale. 
For example, in the case of one generation, to obtain
a light neutrino mass $m\approx 0.1$ eV one should take 
$M_R\approx 10^{14}$ GeV.

Understanding the structure of the RH neutrino sector
is an important theoretical issue.
However, the possibility to investigate such a structure
could seem a too arduous experimental task because of the large mass
of RH neutrinos.
Nevertheless, there are at least two footprints of the seesaw mechanism
at accessible energy scales: the light neutrino mass matrix $m$
and the Baryon Asymmetry of the Universe (BAU). In fact, the seesaw
has a simple and elegant built-in mechanism of production of the
BAU: baryogenesis via leptogenesis \cite{FY}.

The precision in the determination of both the low energy neutrino parameters 
\cite{skatmo,sksolar,sno,snosalt,kamland,chooz} 
and the baryon-to photon ratio $\eta_B$ \cite{WMAP}
increased fast in the last few years. 
The requirement to reproduce these two experimental evidences at the same 
time is a severe test for the type-I seesaw mechanism.
The goal of the present paper is to describe the implications of 
these constraints
for the masses and mixings of RH neutrinos \cite{M4}.

\section{Inverting the seesaw formula}

We will consider the light neutrino mass matrix $m$ in the 
left-handed basis formed by $\nu_e$, $\nu_\mu$ and
$\nu_\tau$. In this {\it flavor basis} the charged lepton mass matrix 
$m_l$ is diagonal. 
The Eq.(\ref{seesaw}) can be rewritten as
\begin{equation}
m = - U_L^\dag m_D^{diag} U_R (M_R^{diag})^{-1} U_R^T m_D^{diag} U_L^* ~,
\label{seeC}
\end{equation}
where $m_D^{diag}\equiv (m_{D1},m_{D2},m_{D3})$ and 
$M_R^{diag}\equiv (M_1,M_2,M_3)$ are real and positive diagonal matrices
with $m_{D1}\le m_{D2}\le m_{D3}$ and $M_1\le M_2\le M_3$.
The unitary matrix $U_L$ describes the mismatch between 
the left-handed rotations diagonalizing $m_l$ and $m_D$.
The unitary matrix $U_R$ describes the mismatch between
the RH rotations diagonalizing $m_D$ and $M_R$.

In the RH basis in which the matrix $U_R$ is absorbed in $M_R$,
inverting the seesaw formula (\ref{seeC}) one obtains 
\begin{equation}
M_R^{-1} \equiv U_R (M_R^{diag})^{-1} U_R^T = 
- \left(\begin{array}{ccc}
\vspace*{0.2cm}
\dfrac{\hat{m}_{ee}}{m_{D1}^2} & \dfrac{\hat{m}_{e\mu}}{m_{D1}m_{D2}} & 
\dfrac{\hat{m}_{e\tau}}{m_{D1}m_{D3}}\\
\vspace*{0.2cm}
\dots & \dfrac{\hat{m}_{\mu\mu}}{m_{D2}^2} & \dfrac{\hat{m}_{\mu\tau}}{m_{D2}m_{D3}}\\
\vspace*{0.2cm}
\dots & \dots & \dfrac{\hat{m}_{\tau\tau}}{m_{D3}^2} 
\end{array}\right) 
~,
\label{mr-1}
\end{equation}
where 
\begin{equation}
\hat{m}\equiv U_L m U_L^T ~.
\label{tilm}
\end{equation}
For simplicity, we denote the entries of $\hat{m}$ with 
$e,~\mu,~\tau$ indexes, even though $\hat{m}$ is the light 
neutrino mass matrix 
in a basis rotated with respect to the flavor basis.

Even if the matrix $m$ were completely known from experiments
(see discussion in section \ref{lowm}), 
one cannot infer the masses of RH neutrinos unless
some assumption is made on the Dirac mass matrix $m_D$.
In this paper we will analyze only the case of 
hierarchical mass spectrum for the neutrino Dirac masses:
\begin{equation}
m_{D1}\ll m_{D2}\ll m_{D3} ~.
\label{hie}
\end{equation}
This choice is motivated by the assumption that
$m_D$ is analogue to the Dirac mass matrices 
of quarks or charged leptons. This is the case in many 
theories with quark-lepton symmetry (for example, 
in minimal $SO(10)$ one has $m_D=m_u$ at GUT scale).
In other words, we assume that the hierarchy among different generation of
Yukawa couplings is a property which holds also in the neutrino sector.
Other studies of the seesaw mechanism and leptogenesis with hierarchical
neutrino Dirac masses can be found in \cite{fatra,orloff,bra1,falco,falast,
velase,wernerH}.

To quantify the strength of the hierarchy that one should expect 
among $m_{Di}$,
we report in Table \ref{table} the approximate 
values of charged lepton and quark masses 
at the renormalization scale $10^{9}$ GeV (SM case \cite{qm}). 
In fact, we are interested in the value of neutrino Dirac masses 
at the scale of RH neutrino masses.
For numerical estimates, we will use for $m_{Di}$ the values
of up-quark masses (the case of minimal $SO(10)$) given in Table \ref{table}.

\begin{table}[h]
\begin{center}
\begin{tabular}{|c||c|c|c|}
\hline
& 1st generation & 2nd generation & 3rd generation\\
\hline\hline
up-type quarks & $1$ MeV & $400$ MeV & $100$ GeV \\
\hline
down-type quarks & $3$ MeV & $50$ MeV & $2$ GeV \\
\hline
charged leptons & $0.5$ MeV & $100$ MeV & $2$ GeV \\
\hline
neutrinos & $m_{D1}$ & $m_{D2}$ & $m_{D3}$\\
\hline
\end{tabular}
\end{center}
\caption{Approximate values of SM quark and charged lepton masses 
renormalized at $10^9$ GeV \cite{qm}. \label{table}}
\end{table}

\section{Basic features of low energy neutrino mass matrix
\label{lowm}}

The r.h.s. of Eq.(\ref{mr-1}) depends on $m$, $U_L$ and $m_{Di}$.
At present, we have direct experimental access only to the
Majorana mass matrix of light neutrinos, $m$. It can be written 
in terms of the observables as 
\begin{equation}
m = U_{PMNS}^*m^{diag}U_{PMNS}^{\dag}~,
\label{matr}
\end{equation}
where $m^{diag} \equiv diag(m_1,~m_2,~m_3)$
and 
\begin{equation}
U_{PMNS} = U(\theta_{12},\theta_{23},\theta_{13},\delta)\cdot K_0 ~,~~~~~~~
K_0=diag(e^{i\rho},1,e^{i\sigma}) ~.
\label{ponm}
\end{equation}
Here $\delta$ is the CP-violating Dirac phase and $\rho$ and $\sigma$ are
the two CP-violating Majorana
phases. 
The matrix $m$ should satisfy a number of experimental constraints.
From the solar, atmospheric, accelerator and reactor neutrino  
experiments we take  
the following  input (at $90\%$ C.L.) 
\cite{skatmo,sksolar,sno,snosalt,kamland,chooz}: 
\begin{equation}
\begin{array}{l}
\Delta m^2_{sol}\equiv\Delta m^2_{12} =
\left( 7.1^{~ +1.9}_{~ -1.1} \right) \cdot 10^{-5}~{\rm eV}^2\;,~~~~~~
\tan^2 \theta_{12}=  0.40^{ + 0.12}_{ - 0.09} ;\\ 
\Delta m^2_{atm}\equiv\Delta m^2_{23}=
\left(2.0^{~ +1.1}_{~ -0.7} \right) \cdot 10^{-3} ~{\rm eV}^2\;,~~~~~~
\tan \theta_{23} = 1 ^{~+ 0.35}_{~- 0.25}; \\
\sin\theta_{13} \lesssim 0.2\;.  
\end{array}   
\label{data}
\end{equation}

A significant freedom in the structure of the mass matrix $m$
still exists due to the unknown absolute mass scale $m_1$ and CP-violating 
phases $\rho$ and $\sigma$ \cite{M1,M2}. 
In spite of this freedom, a generic feature of the  
matrix $m$ emerges: all its elements are of the same order (within a factor of 
10 or so of each other), except in some special cases. The reason for this 
is twofold: 
\begin{itemize}
\item[(a)] a relatively weak hierarchy between the mass eigenvalues:
\begin{equation}
\frac{m_2}{m_3} \geq R\equiv
\sqrt{\frac{\Delta m^2_{sol}}{\Delta m^2_{atm}}}
= 0.19 ^{+0.07} _{-0.05} ~;
\end{equation}
\item[(b)] two large mixing angles $\theta_{12}$ and $\theta_{23}$.
\end{itemize}
A strong hierarchy among certain elements of $m$ can be realized for
specific values of $m_1$, $\rho$ and $\sigma$.
For example,
in the case of normal mass hierarchy ($m_1\ll m_2, m_3$),
the $e$-row elements of $m$ (that is $m_{ee}$, $m_{e\mu}$ and $m_{e\tau}$) are
smaller than $\mu\tau$-block ones by a factor of order $R$ or
$s_{13}$. 
An $e$-row element can vanish for specific values of $\sigma$.

Let us consider how these features of $m$ reflect on $\hat{m}$.
The neutrino masses $m_i$ are of course 
basis-independent. On the contrary, 
the mixing angles and the phases in $\hat{m}$ take values
$\hat{\theta}_{ij},~\hat{\delta},~\hat{\rho},~\hat{\sigma}$
different with respect to flavor basis.
As a consequence, the condition (a) applies both for $m$ and $\hat{m}$,
while (b) can be no longer 
valid in the rotated matrix $\hat{m}$, 
if large mixing angles are contained in $U_L$ (see Eq.(\ref{tilm})).
However, since the matrix $U_L$ is the leptonic analogue of the quark 
CKM mixing matrix, by analogy one can assume that $U_L$ 
is close to the unit matrix (does not contain large mixing angles).
Therefore one expects that, as in the case of $m$, 
all the elements of $\hat{m}$ are
of the same order, apart from special cases.

Even if there are large mixings in $U_L$, the vanishing of some
elements in $\hat{m}$ requires special cancellations. For example, in the
case of normal hierarchy, the $e$-row elements 
($\hat{m}_{ee}$, $\hat{m}_{e\mu}$, $\hat{m}_{e\tau}$)
vanish if both $\hat{\theta}_{12}$ and $\hat{\theta}_{13}$ 
are vanishing. This requires
that $U_L$ in Eq.(\ref{tilm}) contains a large $1-2$ mixing angle
which cancels exactly the large solar mixing in $m$.
In the following we assume that these cancellations do not take place.
We will further comment on the case of large left-handed Dirac-type mixing
in section \ref{stab}.

Using the low energy data we can study, in particular, 
the condition $\hat{m}_{ee}\rightarrow 0$,
which will turn out to be crucial in the following discussion.
If $U_L=\mathbb{1}$, $\hat{m}_{ee}=m_{ee}$.
Using Eq.(\ref{matr}) and the standard parameterization for the matrix $U$
one obtains
$$
m_{ee} = \cos^2\theta_{13} (m_1 e^{-2i\rho} \cos^2\theta_{12} + 
m_2 \sin^2\theta_{12}) + \sin^2\theta_{13} e^{2i(\delta-\sigma)} m_3 ~.
$$
The condition  $m_{ee}\rightarrow 0$ is satisfied for 
\begin{equation}
\tan^2\theta_{13}\approx -\dfrac{m_1 e^{-2i\rho} \cos^2\theta_{12} + 
m_2 \sin^2\theta_{12}} {e^{2i(\delta-\sigma)} m_3}~.
\label{zeroee}
\end{equation}
In the limit $\sin\theta_{13}=0$, 
Eq.(\ref{zeroee}) implies $\rho\approx \pi/2$ and
\begin{equation}
m_1\approx \dfrac{\tan^2\theta_{12}\sqrt{\Delta m^2_{sol}}}
{\sqrt{1-\tan^4\theta_{12}}}
\approx (3-4)\cdot 10^{-3} {\rm eV} ~.
\label{m1mee}
\end{equation}
This corresponds to normal mass hierarchy.
Non-zero $\sin\theta_{13}$ shifts the value of $m_1$ corresponding to
$m_{ee}\rightarrow 0$. Taking into
account the present upper bound on $\sin\theta_{13}$ (Eq.(\ref{data})), 
we find that the relation
(\ref{zeroee}) can be satisfied for $m_1\lesssim 0.02$ eV.
Notice that larger values of $m_1$ are forbidden because $\theta_{12}$ is 
far from the maximal value $\theta_{12}^{max}\equiv\pi/4$.

If $U_L\neq\mathbb{1}$, the condition $\hat{m}_{ee}\rightarrow 0$ 
is satisfied if the
angles $\hat{\theta}_{ij}$ and the phases 
$\hat{\delta},~\hat{\rho}$ and $\hat{\sigma}$ fulfill Eq.(\ref{zeroee}).
This is possible also for  mass spectra different from 
normal hierarchy. 
In this case $m_2\approx m_1 \gg \sqrt{\Delta m^2_{sol}}$ 
and Eq.(\ref{zeroee}) can be satisfied, e.g., for 
$\hat{\theta}_{13}=0$, $\hat{\theta}_{12}=\pi/4$ and $\hat{\rho}=\pi/2$.
These values of mixing angles can be obtained for $U_L\approx U_{CKM}$
\cite{giuta}.

Notice that the neutrinoless $2\beta$ decay experiments \cite{H-M,IGEX} 
restrict 
the $ee$-element of the matrix $m$:
$$
|m_{ee}| < (0.35 \div 1.3) ~{\rm eV}~ \qquad(90\% ~{\rm C.L.})\,.
$$
If future experiments will find a positive signal, this will imply
that $m_{ee}$ is not very small
(unless non-standard mechanism contribute to neutrinoless
$2\beta$-decay rate \cite{verga}). In this case the condition $\hat{m}_{ee}
\rightarrow 0$ could be satisfied only for non-negligible rotations in $U_L$.

\section{Mass spectrum and mixing of RH neutrinos \label{mami}}

Let us compute the eigenvalues and the mixing angles of the matrix $M_R^{-1}$
defined in Eq.(\ref{mr-1}).

\underline{\bf Generic case.}\\ 
The denominators in the r.h.s. of Eq.(\ref{mr-1}) are strongly hierarchical.
As a consequence, unless a special suppression of $\hat{m}_{ee}$ takes place,
the largest eigenvalue of $M_R^{-1}$ is given, to a very good approximation, by
the dominant $11$-element:
\begin{equation}
M_1\approx \dfrac{1}{|(M_R^{-1})_{11}|} = \dfrac{m_{D1}^2}{|\hat{m}_{ee}|} ~.
\label{M1}
\end{equation}
The second largest eigenvalue of $M_R^{-1}$ can be obtained from the dominant
$(12)$-block of the matrix (\ref{mr-1}), just by dividing its determinant
by $(M_R^{-1})_{11}$. The mass $M_2$ is then the inverse of this eigenvalue:
\begin{equation}
M_2 \approx \dfrac{|(M_R^{-1})_{11}|}
{|(M_R^{-1})_{11}(M_R^{-1})_{22}-(M_R^{-1})_{12}^2|} =
\dfrac{m_{D2}^2 |\hat{m}_{ee}|}{|d_{12}|}
~,
\label{M2}
\end{equation}
where 
$$
d_{12}\equiv \hat{m}_{ee} \hat{m}_{\mu\mu} - \hat{m}_{e\mu}^2 ~.
$$
The Eq.(\ref{M2}) is reliable as far as the subdeterminant $d_{12}$ is
not vanishing.
The smallest eigenvalue of $M_R^{-1}$ can be found from the condition 
$$
(m_{D1}m_{D2}m_{D3})^2 = m_1m_2m_3M_1M_2M_3 ~
$$
which is obtained by taking the determinants of both sides of
Eq.(\ref{seesaw}). This yields
\begin{equation}
M_3 \approx \dfrac
{m_{D3}^2 |d_{12}|} {m_1m_2m_3}~.
\label{M3}
\end{equation}
Thus, in the generic case the RH neutrinos have a very strong mass hierarchy: 
$M_1 \propto m_{D1}^2$,  $M_2 \propto m_{D2}^2$,  $M_3 \propto m_{D3}^2$.
Assuming $U_L\approx \mathbb{1}$, the numerical values of $M_i$ are functions
of low energy data and $m_{Di}$ only. One finds \cite{M4}
$$
\begin{array}{l}
M_1\approx (1 - 500) {\rm~TeV} \left(\dfrac{m_{D1}}{1{\rm~MeV}}\right)^2~,\\
M_2\approx (0.2 - 6)\cdot 10^9 {\rm~GeV} 
\left(\dfrac{m_{D2}}{400{\rm~MeV}}\right)^2~,\\
M_3\gtrsim 5\cdot 10^{12} {\rm~GeV} 
\left(\dfrac{m_{D3}}{100{\rm~GeV}}\right)^2~,
\end{array}
$$
where the given ranges reflect our ignorance on the type of mass spectrum 
of light neutrinos.

The matrix $M_R$ is diagonalized, to a high accuracy, by 
\begin{equation}
U_R \approx \left(
\begin{array}{ccc}
\vspace*{0.2cm}
1 & -\left(\dfrac{\hat{m}_{e\mu}}{\hat{m}_{ee}}\right)^*\dfrac{m_{D1}}{m_{D2}} &
\left(\dfrac{d_{23}}{d_{12}}\right)^*\dfrac{m_{D1}}{m_{D3}} \\
\vspace*{0.2cm}
\left(\dfrac{\hat{m}_{e\mu}}{\hat{m}_{ee}}\right)\dfrac{m_{D1}}{m_{D2}} & 1 &
-\left(\dfrac{d_{13}}{d_{12}}\right)^*\dfrac{m_{D2}}{m_{D3}} \\
\vspace*{0.2cm}
\left(\dfrac{\hat{m}_{e\tau}}{\hat{m}_{ee}}\right)\dfrac{m_{D1}}{m_{D3}} &
\left(\dfrac{d_{13}}{d_{12}}\right)\dfrac{m_{D2}}{m_{D3}} & 1 
\end{array}\right) \!\cdot K ~,
\label{rimi}
\end{equation}
where
$$
d_{23}\equiv \hat{m}_{e\mu} \hat{m}_{\mu\tau}
-\hat{m}_{\mu\mu}\hat{m}_{e\tau}~,~~~~~
d_{13}\equiv \hat{m}_{ee}\hat{m}_{\mu\tau}-\hat{m}_{e\mu}\hat{m}_{e\tau}
$$
and
\begin{equation}
K=diag(e^{-i\phi_1/2},e^{-i\phi_2/2},e^{-i\phi_3/2}) ~.
\label{Phi}
\end{equation}
The differences between the Majorana phases $\phi_i$ of RH neutrinos have 
physical meaning, analogously to the case of light neutrinos 
(see Eq.(\ref{ponm})).
As can be seen from Eq.(\ref{rimi}), all the three RH mixing angles
are very small in the generic case 
($\lesssim m_{D1}/m_{D2}$, $m_{D2}/m_{D3}$). 
If also the left-handed  mixing angles in $U_L$ are small, still 
one can obtain a strong mixing in the low-energy sector. 
This is the  so-called ``seesaw enhancement'' of the leptonic mixing 
\cite{enha}. 
The reason for this enhancement can be readily understood. Indeed, small
mixing in $m_D$ and $M_R$ is related to the hierarchical structures of these 
matrices; however, in the seesaw formula (\ref{seesaw}) these hierarchies
act in the opposite directions and largely compensate each other, leading
to a ``quasi-democratic'' $m$ and thus to large mixing in the low-energy 
sector.

\underline{\bf  Special case $\hat{m}_{ee}\rightarrow 0$.}\\
When
\begin{equation}
|\hat{m}_{ee}| \ll \dfrac{m_{D1}}{m_{D2}} |\hat{m}_{e\mu}| ~,
\label{cond}
\end{equation}
the $12$-block of $M_R^{-1}$ in Eq.(\ref{mr-1}) is 
dominated by the off-diagonal entries and, to a good approximation, the 
two lightest RH neutrinos have opposite CP-parity and equal masses: 
\begin{equation}
M_1 \approx  M_2 \approx \dfrac{1}{|(M_R^{-1})_{12}|}\approx
\dfrac{m_{D1} m_{D2}}{|\hat{m}_{e\mu}|} 
~,~~~~~~
M_3\approx \dfrac{m_{D3}^2 |\hat{m}_{e\mu}|^2}{m_1m_2m_3}
~.
\label{S1}
\end{equation} 
Notice that $M_1$ is increased by a factor $\sim m_{D2}/m_{D1}$ with respect to
the generic case (Eq.(\ref{M1})).
Assuming $U_L\approx\mathbb{1}$, one obtains \cite{M4}
\begin{eqnarray}
M_{1,2} &\approx &  
9\cdot 10^{7} ~{\rm GeV} \left(\frac{m_{D1}}{1 ~{\rm MeV}}\right)\!
\left(\frac{m_{D2}}{400 ~{\rm MeV}}\right) \label{M12S1}\,, \nonumber\\
M_3 & \approx & 
10^{14} ~{\rm GeV} \left(\frac{m_{D3}}{100 ~{\rm GeV}}\right)^2 
\label{M3S1} ~. \nonumber
\end{eqnarray} 
These predictions are more precise than in the generic case, since
the light neutrino mass spectrum is fixed by the
condition $m_{ee}\rightarrow 0$ (see Eq.(\ref{m1mee})).

The RH $1-2$ mixing is nearly maximal while the other mixing 
angles remain very small:
\begin{equation}
U_R \approx \left(
\begin{array}{ccc}
\dfrac{1}{\sqrt{2}} & \dfrac{1}{\sqrt{2}} & 
-\left(\dfrac{d_{23}}{\hat{m}_{e\mu}^2}\right)^*
\dfrac{m_{D1}}{m_{D3}} 
 \\\\ 
-\dfrac{1}{\sqrt{2}}  & \dfrac{1}{\sqrt{2}} &
-\left(\dfrac{\hat{m}_{e\tau}}{\hat{m}_{e\mu}}\right)^*\dfrac{m_{D2}}{m_{D3}} \\\\
-\dfrac{\hat{m}_{e\tau}}{\sqrt{2}\hat{m}_{e\mu}}\dfrac{m_{D2}}{m_{D3}}  &
\dfrac{\hat{m}_{e\tau}}{\sqrt{2}\hat{m}_{e\mu}}\dfrac{m_{D2}}{m_{D3}} & 1 \\
\end{array}\right)\!\cdot K ~.
\label{rimiS1}
\end{equation}
The matrix of phases $K$ is given in Eq. (\ref{Phi}) and one has, 
in particular, $\phi_1-\phi_2\approx \pi$. 
Thus, the RH neutrinos $N_1$ and $N_2$ are 
quasi-degenerate, have nearly opposite CP-parities and almost maximal mixing
($1-2$ level crossing).
The third RH neutrino $N_3$ is much heavier and weakly mixed with the 
first two.

\underline{\bf Special case $d_{12} \rightarrow 0$.}\\
Let us consider the case in which the $(11)$-element of the matrix $M_R^{-1}$ 
in Eq.(\ref{mr-1}) is still the dominant one (as in the generic case), but the 
$(12)$-subdeterminant of $M_R^{-1}$ is very small. Then 
$(M_R)_{33}$, which is proportional to this subdeterminant, is suppressed. 
The condition $(M_R)_{33}\ll(M_R)_{23}$ can be written as  
\begin{equation}
|d_{12}|\ll \dfrac{m_{D2}}{m_{D3}} |d_{13}| ~.
\label{small12}
\end{equation}
In this case $M_1$ is still given by Eq.(\ref{M1}),
but the $(23)$-block of $M_R$  
is dominated by its off-diagonal entry. This yields  
\begin{equation}
M_2 \approx M_3 \approx |(M_R)_{23}|= \dfrac{m_{D2} m_{D3}}{m_1m_2m_3} 
|d_{13}| ~.
\label{M23}
\end{equation}
The matrix $U_R$ is similar to the one in Eq.(\ref{rimiS1}) but with
maximal mixing in the $2-3$ sector \cite{M4}.

\underline{\bf Special case $\hat{m}_{ee}\rightarrow 0$ $\&$ 
$d_{12}\rightarrow 0$.}\\
Consider the case when 
\begin{equation}
|\hat{m}_{ee}|\ll\dfrac{m_{D1}}{m_{D3}}|\hat{m}_{e\tau}|~,~~~~~
|\hat{m}_{e\mu}|\ll\dfrac{m_{D2}}{m_{D3}}|\hat{m}_{e\tau}|,
~~\dfrac{m_{D1}}{m_{D2}}|\hat{m}_{\mu\mu}|~.
\label{condi3}
\end{equation}
Then both $\hat{m}_{ee}$ and $d_{12}$ are vanishing.
The $(13)$- and $(22)$-elements of $M_R^{-1}$ are 
the dominant ones (see Eq.(\ref{mr-1})).
Two RH neutrinos form a quasi-degenerate pair with almost 
maximal mixing and opposite CP-parities
and the third neutrino has small mixing with the other two (of order 
$m_{D1}/m_{D2}$ or $m_{D2}/m_{D3}$).
The masses of these doublet and singlet states are given by
\begin{equation}
M_d\approx \dfrac{1}{|(M_R^{-1})_{13}|}\approx  
\dfrac{m_{D1} m_{D3}}{|\hat{m}_{e\tau}|} ~,~~~~~
M_s \approx \dfrac{1}{|(M_R^{-1})_{22}|}\approx 
\dfrac{m_{D2}^2}{|\hat{m}_{\mu\mu}|} ~.
\label{S3}
\end{equation}
Since $m_{D1} m_{D3} \sim m_{D2}^2$, all the three masses are 
of the same order ($\sim 10^{10}$ GeV). The explicit form of $U_R$ 
for this case can be found in \cite{M4}.\\

\begin{figure}[th]
\begin{center}
\epsfig{file=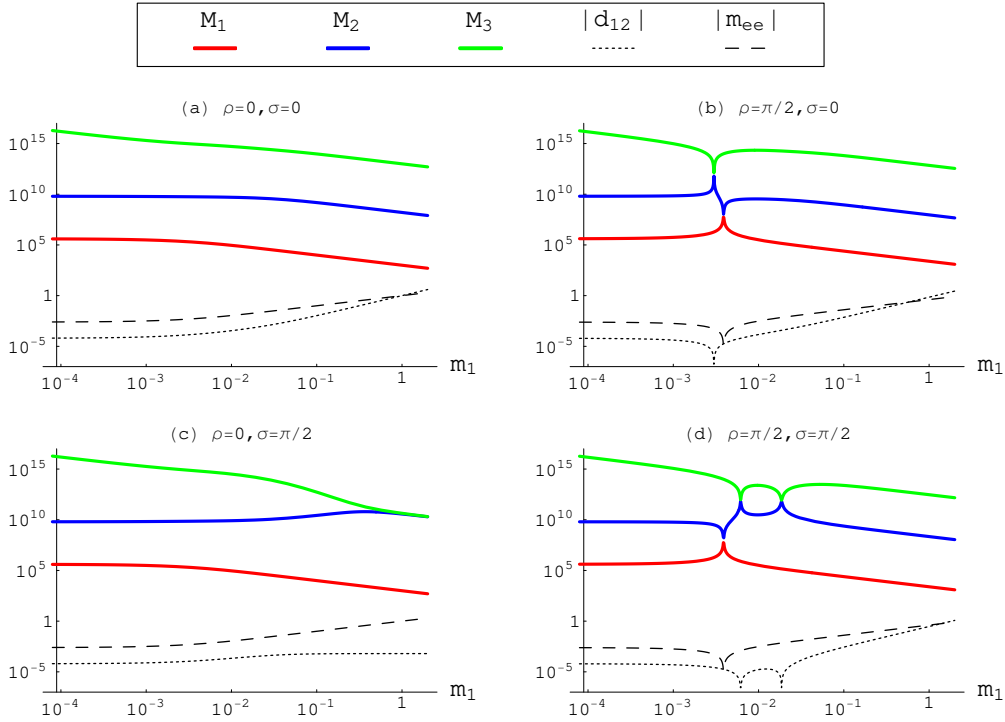,width=420pt}
\caption{
The masses of RH neutrinos $M_i$ in GeV as functions of the light neutrino
mass $m_1$ in eV (solid thick lines), 
for different values of the Majorana phases of light 
neutrinos, $\rho$ and $\sigma$. We assume normal mass ordering; 
$U_L=\mathbb{1}$;
$s_{13}=0$; best fit values of solar and atmospheric mixing angles and 
mass squared differences (Eq.(\ref{data})); values of $m_{Di}$ given by
the up-type quark masses in Table \ref{table}. 
Also shown are $|d_{12}|\equiv |m_{ee} m_{\mu\mu}-m_{e\mu}^2|$ in eV$^2$ 
(dotted thin line) and $|m_{ee}|$ in eV (dashed thin line) as functions of 
$m_1$.}
\label{fig}
\end{center}
\end{figure}

Notice that in all the three special cases, the mass-degeneracy of two
RH neutrinos is associated with almost maximal mixing between them and
opposite relative CP-parity. In fact, when mass hierarchies in 
$m_D$ and $M_R$ do not compensate,
a strongly off-diagonal structure of $M_R$ is necessary for the seesaw
enhancement of lepton mixing \cite{enha}.

The features of RH neutrino mass spectrum can be seen in
Fig.\ref{fig}, where, assuming $\hat{m}=m$, we show the dependence 
of the RH neutrino masses 
on the lightest mass $m_1$, 
for different values of the Majorana phases of the light neutrinos $\rho$ 
and $\sigma$. 
One sees immediately that the crossing points where $M_1\approx M_2$ 
correspond to $m_{ee}\rightarrow 0$.
The other possible level crossing ($M_2\approx M_3$) is realized 
when $d_{12} \rightarrow 0$.

\section{Baryogenesis via leptogenesis \label{BVL}}

Let us consider the constraints on the seesaw parameters coming from the 
requirement of successful thermal leptogenesis. We assume that  
a lepton asymmetry $\epsilon_i$ is generated by the CP-violating 
out-of-equilibrium decays of the RH neutrino $N_i$ 
in the early Universe \cite{FY}:
$$
\epsilon_i=\dfrac
{\Gamma(N_i\rightarrow L\phi)-
\Gamma(N_i\rightarrow \bar{L}\bar{\phi})}
{\Gamma(N_i\rightarrow L\phi)+
\Gamma(N_i\rightarrow \bar{L}\bar{\phi})}
~,
$$
where $L$ and $\phi$ are the SM lepton and Higgs doublets.
The lepton asymmetry is then 
converted to a baryon asymmetry through the sphaleron processes \cite{KRS},
thus explaining the baryon asymmetry
of the Universe. We will use the recent experimental value of the 
baryon-to-photon ratio 
\cite{WMAP},
\begin{equation}
\eta_B = (6.5 \pm ^{0.4} _{0.3})\cdot 10^{-10} ~.
\label{etaB}
\end{equation}

The lepton number asymmetry $\epsilon_i$
can be written as
\cite{luty,FPS,plum,CRV,buch1}:
\begin{equation}
\epsilon_i = \dfrac{1}{8 \pi} \sum_{k\neq i} 
f\left(\dfrac{M_k^2}{M_i^2}\right) 
\dfrac{{\rm Im}[(h^\dag h)_{ik}^2]}{(h^\dag h)_{ii}}~. 
\label{epsi}
\end{equation}
Here $h$ is the matrix of neutrino Yukawa couplings in the basis where
$M_R$ is diagonal with real and positive eigenvalues. 
Using the relation $h \equiv m_D/v = ( U_L^\dag m_D^{diag} U_R )/v$ 
(where $v=174$ GeV is the electroweak VEV)
we can write
\begin{equation}
h^\dag h = \dfrac{1}{v^2} U_R^\dag (m_D^{diag})^2 U_R ~. 
\label{hh}
\end{equation}
In the SM the function $f$ in Eq. (\ref{epsi}) is given by
\begin{equation}
f(x)=\sqrt{x}\left[
\dfrac{2-x}{1-x}-(1+x)\log\left(\dfrac{1+x}{x}\right)
\right] ~.
\label{f}
\end{equation}
This expression is valid for $|M_i-M_j| \gg \Gamma_i+\Gamma_j$, where 
$\Gamma_i$ is the decay width of the $i$th RH neutrino, given at tree level by
$$
\Gamma_i = \dfrac{(h^\dag h)_{ii}}{8\pi} M_i ~.
$$
In the limit of the quasi-degenerate neutrinos ($x = M_j^2/M_i^2 
\rightarrow 1$), one formally obtains from (\ref{f})
\begin{equation}
f(x) \approx \frac{1}{1 - x} \approx \frac{M_i}{2(M_i-M_j)} 
\rightarrow \infty~.
\label{fdeg}
\end{equation}
However, in reality the enhancement of the asymmetry  
is limited by the decay widths $\Gamma_i$ and 
is maximized when $|M_i-M_j| \sim \Gamma_i+\Gamma_j$ \cite{pila,pilaU,ERY}.

The baryon-to-photon ratio can be written as \cite{BDP1}
$$
\eta_B \simeq 0.01 \sum_i \epsilon_i \cdot \kappa_i~,
$$
where the factors $\kappa_i$ describe the washout of the 
produced lepton asymmetry 
$\epsilon_i$ due to various lepton number violating processes. 
In the domain of the parameter space which is of interest to us, 
they depend mainly on the effective mass parameters 
\begin{equation}
\tilde{m}_i\equiv\dfrac{v^2(h^\dag h)_{ii}}{M_i} =
\dfrac{[U_R^\dag (m_D^{diag})^2 U_R]_{ii}}{M_i}~.
\label{tildem}
\end{equation}
For $10^{-2}~{\rm eV} <\tilde{m}_1<10^3~{\rm eV}$, the washout factor 
$\kappa_1$ can be well approximated by \cite{KT} 
\begin{equation}
\kappa_1 (\tilde{m}_1)\simeq 0.3 \left(\dfrac{10^{-3}~{\rm eV}}{\tilde{m}_1}
\right)
\left(\log\dfrac{\tilde{m}_1}{10^{-3}~{\rm eV}}\right)^{-0.6}~.
\label{k1}
\end{equation}
When $M_1 \ll M_{2,3}$, only the decays of the lightest RH neutrino 
$N_1$ are relevant for producing the baryon asymmetry $\eta_B$, since the 
lepton asymmetry generated in the decays of the heavier RH neutrinos 
is washed out by the $L$-violating processes involving $N_1$'s, which are 
very abundant at high temperatures $T\sim M_{2,3}$. At the same time,
at $T\sim M_1$ the heavier neutrinos $N_2$ and $N_3$ have already
decayed and so cannot wash out the asymmetry produced in the decays of
$N_1$.

For a recent systematic study of thermal leptogenesis with a detailed 
analysis of washout effects see \cite{totale}.

\section{A unique structure for successful thermal lepto\-genesis}

Let us compute the value of $\eta_B$ generated through the decays of
RH neutrinos in the different cases discussed in section \ref{mami}.
 
\underline{\bf Generic case.}\\
From Eqs.(\ref{tildem}), (\ref{M1}) and (\ref{rimi}) we get
\begin{equation}
\tilde{m}_1\approx\dfrac{|\hat{m}_{ee}|^2+|\hat{m}_{e\mu}|^2+
|\hat{m}_{e\tau}|^2}{|\hat{m}_{ee}|}
~.
\label{tildem1}
\end{equation}
Assuming $U_L\approx\mathbb{1}$ and using low energy data,
it turns out \cite{M4} that $\tilde{m}_1 \gtrsim \sqrt{\Delta m^2_{sol}}$,
so that Eq.(\ref{k1}) implies $\kappa_1 \lesssim  0.02$.
From Eqs.(\ref{epsi})-(\ref{f}) and (\ref{M1})-(\ref{rimi}), 
we obtain the following expression for the lepton asymmetry:
$$
\epsilon_1 \approx \dfrac{3}{16 \pi} \dfrac{m_{D1}^2}{v^2}
\cdot I(\hat{m}_{\alpha\beta})~,
$$
where $ I(\hat{m}_{\alpha\beta})$ is an order one function of the elements
of $\hat{m}$.
Then the produced baryon-to-photon ratio is given, up to a factor of order
one, by
$$
\eta_B 
\simeq 0.01 \cdot \epsilon_1 \cdot \kappa_1 \simeq  4\cdot 10^{-16} 
\left(\frac{m_{D1}}{1 ~{\rm MeV}}\right)^2 
\left(\dfrac{\kappa_1}{0.02}\right)~.
$$
To reproduce the observed value of $\eta_B$, 
one would need $m_{D1} \sim 1$ GeV.
Thus, a successful leptogenesis requires $m_{D1} \sim m_{D2}$, 
which contradicts our assumption of a strong hierarchy between the eigenvalues 
of $m_D$ and goes contrary to the simple GUT expectations.   
Therefore, the generic case does not lead to successful leptogenesis.

\underline{\bf Special case $\hat{m}_{ee}\rightarrow 0$.}\\ 
Since $N_1$ and $N_2$ are quasi-degenerate and almost maximally mixed, 
$\epsilon_1$ and $\epsilon_2$ are almost equal.
The dominant contribution to $\epsilon_{1,2}$ is given by 
(see Eqs.(\ref{rimiS1}), (\ref{epsi}), (\ref{hh}) and (\ref{fdeg}))
\begin{equation}
\epsilon_1\approx\epsilon_2\approx\dfrac{1}{16\pi}\dfrac{M_1}{M_1-M_2}
\dfrac{{\rm Im}[(h^\dag h)_{12}^2]}{(h^\dag h)_{11}}\approx
\dfrac{1}{16\pi}\dfrac{m_{D2}^2}{v^2}
\xi ~,
\label{epsigo}
\end{equation}
where
\begin{equation}
\xi=\dfrac{M_1}{M_1-M_2} \sin(\phi_1-\phi_2) ~.
\label{xi}
\end{equation}
The enhancement due to the quasi-degeneracy of $N_1$ and $N_2$ competes
with the suppression due to their almost opposite CP-parities:
$(\phi_1-\phi_2)\approx \pi$.
Starting from Eq.(\ref{mr-1}) and performing a detailed computation 
of the mass splitting and of the deviation of
$\sin(\phi_1-\phi_2)$ from zero, one finds
\begin{equation}
\xi\approx\dfrac{4k\tan\Delta}{(1+k)^2+(1-k)^2\tan^2\Delta}~,
\label{xi1}
\end{equation}
where
\begin{equation}
k \equiv \dfrac{m_{D1}^2|\hat{m}_{\mu\mu}|}{m_{D2}^2 |\hat{m}_{ee}|}~,~~~~~
\Delta\equiv \dfrac 12 \arg\dfrac{\hat{m}_{e\mu}^2}
{\hat{m}_{ee}\hat{m}_{\mu\mu}}~.
\label{k}
\end{equation}
For $|1-k|\ll 1/\tan\Delta$, Eq.(\ref{xi1}) gives  $\xi\approx\tan\Delta $, 
so that for $\Delta\simeq \pi/2$ a significant enhancement of the asymmetries 
$\epsilon_{1,2}$ can be achieved.

Because of almost maximal $1-2$ RH mixing, both $N_1$ and $N_2$ interact with 
the thermal bath mainly via the Yukawa coupling $m_{D2}/v$. This, in contrast
with the generic case, implies $\epsilon_1 \propto m_{D2}^2$ instead of 
$m_{D1}^2$, but also washout effects much stronger. In fact, from 
Eqs.(\ref{tildem}), (\ref{S1}) and (\ref{rimiS1}) we obtain
$$
\tilde{m}_1 \approx \tilde{m}_2 \approx
\dfrac{m_{D2}}{m_{D1}}\dfrac{|\hat{m}_{e\mu}|^2+|\hat{m}_{e\tau}|^2}
{2|\hat{m}_{e\mu}|}
~.
$$
Assuming $U_L\approx\mathbb{1}$ and using for $m_{Di}$ 
the values given in Table \ref{table} for up-type quarks, we find \cite{M4}
$\tilde{m}_1\approx 1.5$ eV and thus Eq.(\ref{k1}) implies 
$ \kappa_1 \approx \kappa_2 \approx  6\cdot 10^{-5}$.

Combining Eqs.(\ref{k1}) and (\ref{epsigo}) and taking into account the
restriction $M_2-M_1\gtrsim \Gamma_1$ (see section \ref{BVL}), 
one finally obtains \cite{M4}
$$
\eta_B \approx 0.01 \cdot 2 \epsilon_1 \kappa_{1}
\lesssim 2\cdot 10^{-8}
\left(\dfrac{400m_{D1}}{m_{D2}}\right)^2
\left[1+0.14\log\left(\dfrac{m_{D2}}{400m_{D1}}\right)\right]^{-0.6}
~.
$$
The value (\ref{etaB}) of $\eta_B$ can be reproduced for
$m_{D1}/m_{D2} \gtrsim 2\cdot 10^{-3}$. 
This corresponds \cite{M4} to a relative splitting 
$(M_2-M_1)/M_1\lesssim 10^{-5}$.

Thus, in spite of strong washout effects, a sufficiently large baryon 
asymmetry can be generated in this special case, 
due to the enhancement related to 
the strong degeneracy of the RH neutrinos. For this to occur, not only 
the level crossing condition ($\hat{m}_{ee}\rightarrow 0$) 
has to be satisfied, 
but also the conditions $\Delta\approx\pi/2$ and $k\approx 1$ should be 
fulfilled, where $\Delta$ and $k$ are defined in Eq.(\ref{k}).  
All these requirements are consistent with the low energy neutrino 
data. We have checked these analytic results by 
precise numerical calculations.

\underline{\bf Other special cases.}\\ 
In the  special case $d_{12}\rightarrow 0$
($M_1\ll M_2\approx M_3$), the produced lepton asymmetry
is dominated by the decays of $N_1$. The RH mixing
angles are larger than in the generic case, but the contributions
to $\epsilon_1$ from diagrams with $N_2$ or
$N_3$ in the loop cancel each other because of their opposite CP-parity
\cite{M4}. The final asymmetry is much smaller than the required value.

In the special case  $\hat{m}_{ee}\rightarrow 0$ $\&$ 
$d_{12}\rightarrow 0$, since the three RH neutrinos have similar masses,
the decays of all three $N_i$'s can contribute
to the produced lepton asymmetry. One finds \cite{M4} that some of the
$\epsilon_i$'s can be large, but correspondingly washout effects are
very strong, because the large $1-3$ RH mixing implies that the pair of
maximally mixed RH neutrinos interacts with the thermal bath via the order
one coupling $m_{D3}/v$. As a consequence, leptogenesis is unsuccessful.

\section{Stability of the result \label{stab}}

In the previous section we have computed the baryon asymmetry produced
through the decays of RH neutrinos, in the framework of type-I seesaw 
mechanism with hierarchical Dirac masses $m_{Di}$
and small left-handed Dirac-type mixing $U_L$. 
We have found that the unique possibility to obtain 
successful thermal leptogenesis is the special case $\hat{m}_{ee}
\rightarrow 0$.
Now we want to give some comments and to make checks 
on the stability of this result.\\

\noindent{\bf 1) Supersymmetry.}

The successful special case works also in the SUSY
version, since the mass scale $M_1\approx M_2\sim 10^8$ GeV can be easily 
smaller than the reheating temperature required to avoid 
gravitino overproduction \cite{grav1,grav2,grav3}.

In the Minimal Supersymmetric SM the electroweak VEV $v$  in
Eq.(\ref{hh}) should be replaced with $v\sin\beta$. However, for
$\tan\beta\gtrsim 3$, this corresponds to a very small rescaling 
of Yukawa couplings. As a consequence, the estimation of the lepton asymmetry
is not significantly modified with respect to the SM case.

In some supersymmetric scenarios, $U_L$ and $m_{Di}$ can be probed 
in lepton flavor 
violating (LFV) decays like $\mu\rightarrow e\gamma$ or $\tau\rightarrow 
\mu\gamma$ \cite{borzu,LMS,ehlr,ER,david}. 
If $U_L=\mathbb{1}$, these decays are strongly suppressed and will 
not be observed. On the contrary,
for $U_L\approx U_{CKM}$ one finds the predicted branching ratios 
to be close to the experimental upper bounds, provided that the 
slepton masses are of the order of $(100\div 200)$ GeV and the neutrino 
Dirac masses $m_{Di}$ take values of the order of quark masses.
Therefore, if future experiments find a signal close to the present
upper bounds, this will not require large rotations in $U_L$.
In any case the successful special case is not constrained by LFV bounds, 
since the condition $\hat{m}_{ee}\rightarrow 0$ can be satisfied both for
$U_L=\mathbb{1}$ and $U_L\neq\mathbb{1}$.\\

\noindent{\bf 2) Flavor symmetries in the RH sector.}

The existence of a pair of degenerate and maximally mixed RH
neutrinos may well be the consequence of some flavor symmetry in the RH
sector, like $SU(2)_H$ under which $N_1$ and $N_2$ transform like a doublet
\cite{MK}. 

It is well known \cite{lelmlt,CM} that a pseudo-Dirac structure of the
light neutrino mass matrix is useful to explain large mixing, especially
in the context of models with inverted mass hierarchy.
We have found that also in the RH neutrino sector 
an approximate  pseudo-Dirac structure of the mass matrix can
have important consequences, both for mixing \cite{enha}
and lepton asymmetry enhancement.\\

\noindent{\bf 3) Radiative corrections.}

Let us discuss the renormalization group equation (RGE) evolution of the 
seesaw mass matrices. 

The structure 
of the effective mass matrix $m$ is stable under the SM 
(or MSSM) radiative corrections \cite{CP,BLP,lindner1,M3}. 
The corrections to its 
matrix elements can be written as
\begin{equation}
\Delta m_{\alpha\beta} \sim (\epsilon_\alpha + \epsilon_\beta) 
m_{\alpha\beta} ~,
\label{corri}
\end{equation}
where $\epsilon_\alpha$ ($\lesssim 10^{-2}$) describes the effect of 
the Yukawa
coupling of the charged lepton $l_\alpha$. The Eq.(\ref{corri}) implies
that both $m_{ee}$ and 
$(m_{ee}m_{\mu\mu}-m_{e\mu}^2)$ receive small corrections 
proportional to themselves: if they are very small at 
the electroweak scale, they remain very small also at the seesaw scale 
(the mass scale of RH neutrinos). Therefore, if $U_L=\mathbb{1}$, 
the level crossing conditions $\hat{m}_{ee}\rightarrow 0$ and $d_{12}
\rightarrow 0$ can be tested with low energy data, at least in principle.
However, non-negligible $U_L$ rotations can lead to the vanishing of
$\hat{m}_{ee}$ and $d_{12}$ at the level crossing energy scale, even though
$m_{ee}$ and $(m_{ee}m_{\mu\mu}-m_{e\mu}^2)$ are not vanishing at low energy.

Between the GUT and the seesaw scales one has to consider the evolution of 
the neutrino Yukawa couplings $h$ and of the Majorana 
mass matrix of RH neutrinos $M_R$ rather than the evolution of the effective 
matrix $m$ \cite{HOS,cein,lindner2,FJ}. 
We assumed that, at the GUT scale, $h$ is related with 
the Yukawa couplings of quarks or charged leptons. 
The evolution of $h$ with decreasing mass scale will not modify the 
hierarchy $m_{D1}\ll m_{D2} \ll m_{D3}$, and its effects can be absorbed into a 
redefinition of our indicative values of $m_{D1,D2,D3}$.

The RGE effects on $M_R$ are due to the neutrino Yukawa couplings; they 
can, in principle, be important in the cases of strongly degenerate RH 
neutrinos. Consider the stability of the structure of $M_R$ in the special 
case that leads to a successful leptogenesis. Recall that in this case the 
$(12)$-sector of RH neutrinos is characterized by $M_{1,2}\sim 10^8$ 
GeV, $(M_2-M_1)/M_1\lesssim 10^{-5}$ and $\Delta\approx \pi/2$, where 
$\Delta$ is defined in Eq.(\ref{k}). The largest correction to the 
$(12)$-block of $M_R$ between $M_{\rm GUT}$ and $M_{1,2}$ is the 
correction to the $22$-element: 
$$
\dfrac{(\Delta M_R)_{22}}{(M_R)_{22}}\sim
\dfrac{m_{D2}^2}{16\pi^2 v^2}
\log\left(\dfrac{M_{\rm GUT}}{10^8{\rm~GeV}}\right) \approx 6\cdot 10^{-7}
\left( \dfrac{m_{D2}}{0.4{\rm~GeV}}\right)^2 ~.
$$
Therefore, the radiative corrections cannot generate a relative splitting 
between $M_1$ and $M_2$ exceeding  $10^{-5}$. Moreover, at one loop level, 
the phases of $(M_R)_{ij}$ have no RGE evolution and so the relation 
$\Delta\approx\pi/2$ is not modified.

It has been recently shown \cite{nobre} that, 
assuming exact degeneracy
of $M_1$ and $M_2$ at the GUT scale, successful leptogenesis can be
realized thanks to the radiatively induced splitting at the scale 
$M_1\approx M_2$.
\\

\noindent{\bf 4) Large left-handed Dirac-type mixing.}

Let us abandon the hypothesis $U_L\approx\mathbb{1}$. If the matrix 
$U_L$ is arbitrary, the connection between the low energy data
and the structure of $M_R$ is weakened. 
This additional freedom relaxes the phenomenological constraints on 
RH neutrinos. In fact, now the unique low energy requirement on the seesaw 
mechanism is to reproduce the light neutrino masses, given by the eigenvalues 
of $\hat{m}$ (Eq.(\ref{tilm})); the correct leptonic mixing matrix 
$U_{PMNS}$ can always be obtained through the proper choice of $U_L$. 
As an example, let us consider the case of non-degenerate RH masses and take 
the following RH mixing matrix:
\begin{equation}
U_R=\left(
\begin{array}{ccc}
\dfrac{1}{\sqrt{2}} & \dfrac{1}{\sqrt{2}} & 0 \\
-\dfrac{1}{\sqrt{2}} & \dfrac{1}{\sqrt{2}} & 0 \\
0 & 0 & 1
\end{array}\right)\cdot K ~.
\label{spU}
\end{equation}
The eigenvalues of the matrix $\hat{m}=-m_D^{diag} M_R^{-1} m_D^{diag}$
are given, approximately, by $m_{D2}^2/(4M_2)$, $m_{D2}^2/
(2M_1)$, $m_{D3}^2/M_3$. Taking $M_1\approx  10^{10}$ GeV$\cdot
(m_{D2}/0.4$ GeV$)^2$, $M_2$ a few times larger and $M_3\approx 2\cdot 
10^{14}$ GeV $\cdot(m_{D3}/100~{\rm GeV})^2$, one can reproduce the solar 
and atmospheric mass squared differences in Eq.(\ref{data}). 
Since $\hat{m}$ is approximately 
diagonal, the solar and atmospheric mixing angles are generated by $U_L$, 
which should have an almost bimaximal form. 

Notice that in this scenario we have
$|\hat{m}_{ee}|\approx (m_{D1}/m_{D2})|\hat{m}_{e\mu}|$. In a sense this
situation is intermediate between the generic case and 
the special case $\hat{m}_{ee}\rightarrow 0$ (compare
with Eq.(\ref{cond})). However it cannot be realized unless large rotations 
are allowed in $U_L$. Notice that, in the SUSY case, these large rotations can
be excluded by future stronger bounds on LFV decays.

Replacing Eq.(\ref{spU}) into Eqs.(\ref{tildem}) and (\ref{epsi}),
it is easy to calculate the washout mass parameter and the asymmetry produced 
in the decays of $N_1$:
$$
\tilde{m}_1=\dfrac{m_{D2}^2}{2M_1}\approx \sqrt{\Delta m^2_{sol}}~,~~~~~
\epsilon_1 \approx \dfrac{3m_{D2}^2}{32\pi v^2}\sin(\phi_2-\phi_1)
\dfrac{M_1}{M_2}~.
$$
Taking $\phi_2-\phi_1\sim \pi/2$ (note that the CP-parities of $N_1$ and 
$N_2$ are not constrained in this case), we get
$$
\eta_B \approx 3\cdot 10^{-11}\dfrac{M_1}{M_2}\left(\dfrac{m_{D2}}
{0.4~{\rm GeV}}\right)^2 ~.
$$
Thus, for a moderate hierarchy between $M_1$ and $M_2$, a value of $m_{D2}$ 
around a few GeV can lead to a successful leptogenesis. 
This example shows that, relaxing the hypothesis $U_L\approx \mathbb{1}$,
it is easier to realize baryogenesis via leptogenesis. In particular, 
the degeneracy of the masses of RH neutrinos $M_i$ is no longer necessary, 
but the hierarchy of $M_i$ should not be as large as it is in the generic 
case.\\

\noindent{\bf 5) Non-thermal leptogenesis.}

Let us comment on the possibility of non-thermal production of
the heavy RH neutrinos \cite{shala,muya,GPRT,FHY,BDPS}, 
that in principle can lead to a successful leptogenesis for values of
the parameters $M_1$ and $\tilde{m}_1$ for which thermal leptogenesis 
does not work.

In fact, it is interesting that also non-thermal leptogenesis is strongly 
constrained in our framework. Consider the generic case.
Since $M_1$ is relatively light ($\lesssim 10^7$ GeV), 
$\epsilon_1$ is very small.
Moreover, as $\tilde{m}_1$ is relatively large 
($\gtrsim \sqrt{\Delta m^2_{sol}}$),
thermal effects  washout (at least partially) the asymmetry
generated in the decays of non-thermally produced RH neutrinos
\cite{snEll}.
As a consequence, even in the non-thermal case, 
the asymmetry generated by $N_1$ turns out to be insufficient and, 
to enhance it, one has
to resort again to the special case $\hat{m}_{ee}\rightarrow 0$. 

It is known, however (see, e.g., \cite{ANT,sensha}), that also the
asymmetries generated by $N_2$ and/or $N_3$ can survive if (1) 
they are produced non-thermally at reheating and (2)
$N_1$ is not in thermal equilibrium at the reheating
temperature $T_{RH}$.
In fact, the asymmetries $\epsilon_{2,3}$ can be large (they are of the order 
of $m_{D2,D3}^2/(16\pi v^2)$ in the generic case and even larger in the
special case $d_{12}\rightarrow 0$: $\epsilon_{2,3}\sim m_{D2}/m_{D3}$).
However, partial thermalization of $N_{2,3}$ and subsequent washout 
can occur after reheating. Moreover, to avoid later cancellation of
$\epsilon_{2,3}$, $N_1$ should not enter into thermal
equilibrium at any temperature $T\lesssim T_{RH}$.

In this case an accurate computation of the final asymmetry 
would require to solve the complete set
Boltzmann equations describing the evolution of the
number densities of all three RH neutrinos and of $B-L$.

\section{Conclusions}

We have analyzed the structure of the RH neutrino sector 
in the framework of type-I seesaw mechanism.
We have found a convenient parameterization in which the mass matrix of
RH neutrinos, $M_R$, is a function of the low energy neutrino data,
the neutrino Dirac-type masses $m_{Di}$ and the 
left-handed Dirac-type mixing matrix $U_L$.
Our analysis is based on the assumptions of hierarchical $m_{Di}$ 
(by analogy with quark masses)
and small mixing in $U_L$ (by analogy with CKM mixing).

The presence of two large mixing angles ($\theta_{12}$ and $\theta_{23}$) 
and the weak mass hierarchy ($\sqrt{\Delta m^2_{sol}/\Delta m^2_{atm}}
\approx 0.2$) in the light neutrino sector lead, in general, to 
a ``quasi-democratic" structure of the mass matrix $m$ in the flavor basis, 
with values of all its elements within one order of magnitude of each other. 
This implies that $M_R$ has a strong (nearly quadratic in $m_{Di}$) 
hierarchy of eigenvalues 
and small mixing. The lightest RH neutrino has a mass $M_1 < 10^6$ GeV.
As a consequence, the predicted $\eta_B$ is of the order of 
$\sim (10^{-16} - 10^{-14})$ 
and the scenario of baryogenesis via leptogenesis  does not work.

We have identified the special cases which correspond to the level crossing 
points, when either 
two or all three masses of RH neutrinos are nearly equal. We have found two 
level crossing conditions:\\ (1) $\hat{m}_{ee}\rightarrow 0$ 
($N_1 - N_2$ crossing);\\
(2) $d_{12}\equiv (\hat{m}_{ee}\hat{m}_{\mu\mu}-\hat{m}_{e\mu}^2) 
\rightarrow 0$ ($N_2 - N_3$ crossing).\\
In the crossing points the mixing of the 
corresponding neutrino states is maximal and their CP-parities are nearly 
opposite. 
The level crossing conditions can be realized in agreement with low-energy 
data, because of the freedom in the choice of light neutrino absolute mass
scale $m_1$ and Majorana phases $\rho$ and $\sigma$, as well as in the choice
of small rotations in $U_L$.

The thermal leptogenesis can be successful only 
in the special case with vanishing $\hat{m}_{ee}$.
It is characterized by $M_1 \approx M_2 \sim 
10^{8}$ GeV, $M_3 \sim  10^{14}$ GeV and $(M_2 - M_1)/M_2 \lesssim  
10^{-5}$. $N_1$ and  $N_2$ are strongly mixed and their mixing with $N_3$ is 
very small. The CP-violating phase $\Delta$ in Eq.(\ref{k}) should be very 
close to $\pi/2$. 
Notice that this unique 
case with a successful leptogenesis is defined very precisely. It has a 
number of characteristic features which can give important hints for model 
building.

We have discussed in detail the stability of this result. It turns out that
the successful scenario works also in the case of SUSY. Moreover it is stable
under radiative corrections. The approximate pseudo-Dirac structure can be
motivated by some flavor symmetry operating in the RH sector.

Can the unique successful special case that we found be ruled out? Since it 
requires a suppression of $\hat{m}_{ee}$, it will be excluded in case of a 
positive signal of neutrinoless $2\beta$-decay with $m_{ee}$ close to the 
heaviest of the light neutrino masses (which could be measured in direct 
neutrino mass search experiments). 
In fact, if $m_{ee}$ takes this ``maximal'' value, than  $\hat{m}_{ee}$ 
cannot vanish unless mixings in $U_L$ are very large.
If the condition  $\hat{m}_{ee}\rightarrow 0$ is not realized, 
one will be left with the following alternatives:

\begin{itemize} 
\item The quark-lepton symmetry is strongly violated: there is no strong
hierarchy of the eigenvalues of $m_D$ ($m_{D1}/m_{D2},m_{D2}/m_{D3}
\gtrsim 10^{-1}$)
and/or the Dirac-type left-handed  
mixing is large ($U_L$ contains the solar and/or atmospheric mixings).
\item Type-I seesaw is not the sole source of neutrino mass; 
the simplest alternative could be 
type-II seesaw \cite{LSW,mose2,SV,wet,gero,masa} 
in which there is an additional contribution 
from an $SU(2)_L$-triplet Higgs. 
Another possibility is that 
the seesaw is not the true mechanism of neutrino mass generation.

\item A mechanism other than the decay of RH neutrinos contributes to 
leptogenesis (for leptogenesis in the presence of an $SU(2)_L$-triplet 
see \cite{lsR,laza,hamsen}) or the Baryon Asymmetry of the Universe 
is generated 
through a different mechanism, which has nothing to do with leptogenesis.

\end{itemize}


\section*{Acknowledgments}

A large part of this paper is based on previous work done in collaboration 
with A. Y. Smirnov and E. K. Akhmedov. 
My participation to XV IFAE, Lecce (Italy), April 2003 has been supported 
by INFN under the program ``Fisica Astroparticellare''. 
I would like to thank the conference organizers and especially D.
Montanino for invitation.
My participations to IV NANP Conference, Dubna (Russia), June 2003 and to 
VIII TAUP Workshop,
Seattle (USA), September 2003 have been supported by the Italian MIUR 
under the program ``Fenomenologia delle Interazioni Fondamentali''. 
I thank the NANP conference organizers for invitation and TAUP organizing
committee for local financial support.
This work has been completed with the support of U.S. Department of Energy
under Grant No. DE-FG03-94ER40837.
I am grateful to F. Feruglio, C. Giunti, Y. A. Kamyshkov, G. Mangano, 
S. Pakvasa, F. Terranova and J. D. Vergados for useful discussions.

\bibliographystyle{JHEP2}
\bibliography{biblio}

\end{document}